# Neuroeconomics of suicide.

Taiki Takahashi[1]

[1] Department of Behavioral Science, Hokkaido University

Corresponding Author: Taiki Takahashi

Email: taikitakahashi@gmail.com

Department of Behavioral Science, Hokkaido University

N.10, W.7, Kita-ku, Sapporo, 060-0810, Japan

TEL: +81-11-706-3057   FAX: +81-11-706-3066

**Acknowledgements:** The research reported in this paper was supported by a grant from the Grant- in-Aid for Scientific Research ("global center of excellence" grant) from the Ministry of Education, Culture, Sports, Science and Technology of Japan.

Summary:

Suicidal behavior is a leading cause of injury and death worldwide. Suicide has been associated with psychiatric illnesses such as depression and schizophrenia, as well as economic uncertainty, and social/cultural factors. This study proposes a neuroeconomic framework of suicide. Neuroeconomic parameters (e.g., risk-attitude, probability weighting, time discounting in intertemporal choice, and loss aversion) are predicted to be related to suicidal behavior. Neurobiological and neuroendocrinological substrates such as serotonin, dopamine, cortisol (HPA axis), nitric oxide, serum cholesterol, epinephrine, norepinephrine, gonadal hormones (e.g., estradiol and progesterone), dehydroepiandrosterone (DHEA) in brain regions such as the orbitofrontal/dorsolateral prefrontal cortex and limbic regions (e.g., the amygdala) may supposedly be related to the neuroeconomic parameters modulating the risk of suicide. The present framework puts foundations for "molecular neuroeconomics" of decision-making processes underlying suicidal behavior.

1. **Introduction:**

As the global annual rate of suicide approximates 15 per 100,000 individuals, it is estimated that one million people worldwide commit suicide each year. Annual rates of non-fatal suicidal behaviour are 10–20 times higher than those of completed suicide (Kerkhof, 2000). Suicidal behavior thus constitutes a major public health problem. This indicates the necessity of effective theoretical frameworks for suicide prevention. During the past 30 years, economists have contributed insights about the economic motivations underlying suicidal behavior. Hamermesh and Soss (1974) formalized a model of the utility maximization decision faced by those contemplating suicide. Their paper, and subsequent work by economists, developed the notion that suicide occurs when the temporally-discounted stream of expected utility (subjective well-being) over a person's lifetime is sufficiently low, perhaps negative, by assuming the subject's decision-making is rational. However, recent studies in behavioral economics and neuroeconomics revealed that people are irrational in terms of economic theory. Therefore, introducing neuroeconomic frameworks is important for a better understanding of suicidal behavior. Also, recent neurobiological studies on suicidal behavior demonstrated that several neurobiological substrates such as serotonin, dopamine, and neuroactive steroid hormones in the brain regions such as the prefrontal cortex and the limbic structures are important determinants of suicidal behavior. Therefore, combining neuroeconomic theory with these neurobiological finding is necessary to establish molecular neurobiological theory of suicidal behavior ("molecular neuroeconomics" of suicide).

This paper is organized in the following manner. In Section 2, I briefly introduce neuroeconomic theory of decision under risk and over time, and economic theory of suicidal behavior. In Section 3, findings in neurobiology regarding the molecular mechanisms of suicidal behavior are briefly reviewed. In Section 4, I proposed several predictions from molecular neuroeconomic theory of suicidal behavior. Some conclusions from this study and future study directions by utilizing the present molecular neuroeconomic theory, and how to test the present theory experimentally in future neuroeconomic studies are also discussed.

2. **Neuroeconomic theory of risky and impulsive decision making**
2.1 Decision under risk

In behavioral economics, decision under risk is formulated with Kahneman-Tversky's prospect theory (Kahneman and Tversky, 1979). In prospect theory, a subjective value of an uncertain outcome x which is received at the probability of p is $v(x) w(p)$, where

$v(x)$ is a value function and $w(p)$ is a probability weighting function. The prospect theory's value function is assumed to be concave for gains, convex for losses, and steeper for losses than for gains. The most popular parametrization of the value function is a power function (Tversky and Kahneman, 1992):

$$v(x) = \begin{cases} x^{\alpha} & (x \geq 0) \\ -\lambda(-x)^{\beta} & (x < 0) \end{cases} \quad (1)$$

where α, β > 0 measure the curvature of the value function for gains and losses, and λ is the coefficient of loss aversion. A recent neuroeconomic study demonstrated that amygdale damage reduced loss aversion (De Martino et al., 2010). The probability weighting function has been parametrized as (Prelec, 1998; Takahashi, 2011):

$$w(p) = \exp[-(-\ln p)^{s}] \quad (2)$$

where s indicates a distortion in subjective probability (note that s=1 corresponds to linear probability weighting), which has been shown to associate with the anterior cingulate activity (Paulus and Frank, 2006)

In economic theory of suicidal behavior, income uncertainty and the costs of maintaining oneself alive have been associated with suicide (Hamermesh and Soss, 1974; Marcotte, 2003; Suzuki, 2008), implying that subjects with strong loss aversion and risk aversion are more likely to commit suicide (Yamawaki et al., 2005). Hence, it can be supposed that smaller α <1 (i.e., strong risk aversion in gain), larger λ (i.e., strong loss aversion) and s may be associated with suicidal behavior. The contribution of parameter β which determines risk-attitude in loss should also be examined, because Jollant et al., (2010) reported that suicidal behavior is associated with disadvantageous decision-making in a gambling task involving loss and a decreased activation in the orbitofrontal cortex (Jollant et al., 2010).

## 2.2 Decision over time (intertemporal choice)

In order to describe impulsivity and irrationality in temporal discounting, the q-exponential time-discount model for delayed rewards has been introduced and experimentally examined (Cajueiro, 2006; Takahashi, 2007; Takahashi et al., 2007a; Takahashi et al., 2008ab; Takahashi, 2009):

$$V_{q+}(D) = V_{q+}(0) / \exp_{q+}(k_{q+}D) = V_{q+}(0) / [1+(1-q+)k_{q+}D]^{1/(1-q+)} \quad (3)$$

where $V_{q+}(D)$ is the subjective value of a reward obtained at delay D, $q+$ is a parameter indicating irrationality in temporal discounting for gain (smaller q+<1 values correspond to more irrational discounting for delayed gains), and $k_{q+}$ is a parameter of impulsivity regarding the reward at delay $D=0$ (q-exponential discount rate at delay D=0). Note that when $q+=0$, equation 3 is the same as a hyperbolic discount function, while $q+\to1$, is the same as an exponential discount function (Cajueiro, 2006; Takahashi, 2009). Kable and Glimcher (2007) reported that $V_{q+}(D)$ is neurally-represented in brain regions such as the striatum.

Furthermore, it is known that delayed gains and losses are distinctly processed in the brain and loss is less steeply temporally-discounted than gains, which is referred to as the "sign effect" (Xu et al., 2009). Therefore, we should prepare the q-exponential discount function for delayed loss:

$$V_{q-}(D) = V_{q-}(0) / \exp_{q-}(k_{q-}D) = V_{q-}(0)/[1+(1-q-)k_{q-}D]^{1/(1-q-)} \qquad (4)$$

where $V_{q-}(D) > 0$ is the (absolute, unsigned) subjective value (subjective magnitude) of a loss at delay D (note that $V_{q-}(D=0) = \lambda V_{q+}(D=0)$, from equation 1), $q-$ is a parameter indicating irrationality in temporal discounting for loss (smaller q-<1 values correspond to more irrational discounting for delayed losses), and $k_{q-}$ is a parameter of impulsivity regarding the loss (i.e., degree of procrastination) at delay $D=0$. Our previous study (Takahashi et al., 2008b) demonstrated that depressive patients are more irrational in intertemporal choices for gain and loss than healthy controls. Also, a recent study reported that impulsive suicide attempters discount delayed rewards steeply (Dombrovski et al., 2011). These findings indicate that irrational suicidal behavior may be associated with large k values and smaller q values across gain and loss. However, a deliberate, forward-looking suicide attempt may be associated with less steeper temporal discounting, because the weight of future loss is significant for a subject with smaller $k_{q-}$. Dombrovski and colleagues (2011) reported non-impulsive (planned) suicide attempts were associated with smaller time-discount rate ("lethal foresight"), supporting this speculation. Tellingly, anhedonia was found to be related to slow temporal discounting behavior (Lempert and Pizzagalli, 2010), also consistent with this prediction.

### 3 Neurobiological substrates of suicidal behavior
### 3.1 Brain regions related to suicidal behavior
Suicide has been associated with psychiatric illnesses such as depression and schizophrenia. Regarding neuroanatomy, abnormalities in white matter and limbic gray matter appear to be associated with the occurrence of suicide attempts (van Heeringen et al., 2011). The orbitofrontal cortex, the cingulate cortex and the inferior parietal lobule may be involved in response inhibition, disadvantageous decision-making in a gambling task, and suicide attempts (Jollant et al., 2010; van Heeringen et al., 2011). Serotonergic functioning in the dorsolateral prefrontal cortex correlates significantly with levels of hopelessness, a strong clinical predictor of suicidal behaviour (van Heeringen et al., 2003). The right amygdala hypertrophy may be a risk factor for suicide attempts (Spoletini et al., 2011). The left anterior limb of the internal capsule was further associated with suicidal behavior (Jia et al., 2010).

### 3.2 Serotonin and dopamine

The serotonergic system has been the most widely investigated neuromodulatory system in studies of suicide attempters and completers (Yanagi et al., 2005). CSF-5HIAA was correlated with depression and suicidal behavior (Asberg et al., 1976), although tryptophan hydroxylase isoform 2 gene is not related to suicide in Japanese population (Mouri et al., 2009). Our previous study (Takahashi et al., 2008b) reported that depression is associated with steep temporal discounting in the near future, in line with the finding. Also, taking SSRI (selective serotonin reuptake inhibitor) has been associated with suicidal behavior (Dudley et al., 2010). Furthermore, mesolimbic dopaminergic transmission has been hypothesized to be reduced in depression and suicide (Bowden et al., 1997). Suda et al. (2009) reported dopamine D2 receptor is associated with suicide. Zhong et al., (2009) reported that serotonin and dopamine determine the shape of the value function in Kahneman-Tversky's prospect theory (equation 1); i.e., risk aversion and loss aversion. Furthermore, both serotonin and dopamine regulate temporal discounting (Takahashi, 2009). Therefore, involvement of serotonergic and dopaminergic systems in suicide should more extensively been studied by employing neuroeconomic frameworks in future studies. Also, nitric oxide is involved in synaptic neurotransmission and associated with suicide (Cui et al., 2010). Iga et al., (2007) demonstrated that BDNF (brain-derived neurotrophic factor) is associated with suicidal behavior. Therefore, the relation between synaptic neurotransmission and suicide should be examined.

### 3.3 Epinephrine and norepinephrine

Adrenaline (epinephrine) is synthesized from tyrosine and phenylalanine in both the adrenal gland and the brain, and is considered both a hormone and a neurotransmitter. Adrenaline is an activator molecule well known to induce several physiological effects (e.g., increased heart rate) and general cognitive enhancement (e.g., increased awareness and attention). Meana et al., (1992) reported an increase in alpha 2-adrenoceptors in the hypothalamus and frontal cortex in suicide victims. Consistent with the report, our previous study (Takahashi et al., 2007b; Takahashi et al., 2010) demonstrated that high level of noradrenergic activity is related to slow temporal discounting, which may result in an increase in the rate of deliberate suicide attempts (i.e., "lethal foresight", Dombrovski et al., 2010).

**3.4 Neuroactive steroid hormones**

The hypothalamic–pituitary–adrenal (HPA) axis is the major biological infrastructure of the human stress system with interconnections between the structures by the hormones CRH, ACTH, and cortisol (a type of glucocorticoids). The abnormalities in HPA may be related to suicide (Kunugi et al., 2004). The expression regulation of the hippocampal glucocorticoid receptor, a key component of the HPA axis, is decreased primarily among suicides (McGowan et al., 2009). Supriyanto et al., (2011) reported a genotype of FK506 binding protein 5 is associated with suicidal behavior. Butterfield et al. (2005) reported that patients who had attempted suicide demonstrated significantly higher DHEA levels than those who had not attempted suicide, indicating the involvement of DHEA in suicidal behavior. Regarding gonadal steroids, low estradiol and progesterone levels were observed to associate with suicidal behavior in women (Baca-Garcia et al., 2010). Our previous studies demonstrated that stress hormones modulate temporal discounting behavior (Takahashi, 2004; Takahashi et al., 2010). In males, testosterone is also associated with temporal discounting (Takahashi et al., 2006). Therefore future studies should investigate how these steroid hormones modulate neuroeconomic parameters, resulting in an exaggerated suicide rate. Furthermore, low serum cholesterol is associated with suicidal behavior (Kunugi et al., 1997).

4. **Implications for neuroeconomics and neurobiolgy of suicidal behavior**

This study is the first to present a possible unified framework for molecular neuroeconomic theory of suicide. Our theoretical considerations lead us to the following predictions: (a) neurobiological substrates/alterations which increase loss aversion ($\lambda$ in equation 1) may increase suicide rates, (b) neurobiological substrates/alterations which increase risk aversion (i.e., decrease $\alpha<1$ in equation 1) may increase suicide rates, (c) neurobiological substrates/alterations which increase time-discount rate (k parameters in equation 3 and 4) and time-inconsistency (i.e., decrease q parameters < 1 in equation 3 and 4) in intertemporal choice may increase impulsive suicide attempts, (d) neurobiological substrates/alterations which decrease time-discount rate (k parameters in equation 3 and 4) may increase deliberate, non-impulsive suicide attempt ("lethal foresight"). Moreover, because suicide attempts are associated with disadvantageous decision-making in the gambling task involving loss (Jollant et al., 2010), it can be predicted that (e) neurobiological substrates/alterations which decrease risk aversion in loss may increase the risk of suicidal behavior. Future neuroeconomic studies should investigate whether these predictions (a)-(e) are correct or not, at the molecular and neuronal level, in addition to the neuroanatomical level. Neuropsychopharmacological treatments may especially be useful for the future investigations. Furthermore, some economic theories incorporated social psychological and cultural factors such as self identity and habit (Becker and Murphy, 2003; Akerlof and Kranton, 2000). These factors, which may be important for a better understanding of suicidal behavior (Sakamoto et al., 2006), should be incorporated into neuroeconomic theory of suicide.